\documentclass[a4paper]{jpconf}
\usepackage{graphicx}
\usepackage{amsmath}
\usepackage{amssymb}
\begin{document}
\title{Direct detection of relic active and sterile neutrinos}
\author{Yu-Feng Li}

\address{Institute of High Energy Physics, Chinese Academy of Sciences, P.O. Box 918, Beijing 100049, China}

\ead{liyufeng@ihep.ac.cn}

\begin{abstract}
Both active and sterile sub-eV neutrinos can form the cosmic neutrino background in the early Universe. We consider the beta-decaying (e.g., $^3$H) and EC-decaying (e.g., $^{163}$Ho) nuclei as the promising targets to capture relic neutrinos in the laboratory. We calculate the capture rates of relic electron neutrinos and antineutrinos against the corresponding beta decay or electron capture (EC) decay backgrounds in the (3+$N_{\rm s}$) flavor mixing scheme, and discuss the future prospect in terms of the PTOLEMY project.
We stress that such direct measurements of hot DM might not be hopeless in the long term.
\end{abstract}

\section{Introduction}
\label{intro}

Both active and sterile sub-eV neutrinos can form the cosmic neutrino background (C$\nu$B) when they were
decoupled from radiation and matter at a temperature of about one MeV and an age of one
second after the Big Bang~\cite{XZ}.
Relic neutrinos played important roles in the evolution of the Universe,
and has been indirectly proved from cosmological data on the
Big Bang nucleosynthesis (BBN), cosmic microwave background (CMB) anisotropies and large-scale structures (LSS)~\cite{PDG}.
Without considering the lepton asymmetries, the temperature and average number density for one species of relic neutrinos
can be expressed as
\begin{equation}
T^{}_\nu = \left(\frac{4}{11}\right)^{1/3}T^{}_\gamma\approx 1.945 \; {\rm K}\,,\quad\quad\quad
n^{}_\nu = \frac{3}{11}n^{}_\gamma\approx 112\;{\rm cm}^{-3}\,.
%     (1)
\end{equation}
As a consequence, one predicts the average three-momentum today for each species of the relic neutrino is
very small, i.e., $\langle p^{}_\nu \rangle = 3T^{}_\nu\approx 5 \times 10^{-4}
 \; {\rm eV}$~\cite{Ringwald}.

Cosmological observations provide indirect evidence for the existence of the C$\nu$B, however,
it is a great challenge to the present experimental techniques for the direct detection in a laboratory experiment.
Among several detection possibilities~\cite{Ringwald}, the most promising one seems to be the neutrino capture experiment
using radioactive $\beta$-decaying nuclei~\cite{Weinberg,Irvine,Cocco,Vogel,Blennow,Kaboth,LLX,LX11,Liao,Long14}.
The PTOLEMY project~\cite{PTOLEMY} aims to detect the C$\nu$B using 100 grams
of $^3$H as the capture target. Other interesting methods include the electron-capture (EC) decaying
nuclei~\cite{Cocco2,Lusignoli,LX11EC,LXJCAP}, the annihilation of extremely high-energy cosmic neutrinos %(EHEC$\nu$s)
at the $Z$-resonance~\cite{Weiler82,Eberle04,Barenboim04}, and the atomic de-excitation method~\cite{atomic}.

The remaining parts of this work are organized as follows. In Sec.~2 we introduce methods of relic neutrino captures on the
beta-decaying and EC-decaying nuclei, and calculate the rates and energy spectra of neutrino capture rates.
Sec.~3 is devoted to the flavor effects of relic neutrino captures including the neutrino mass hierarchy and presence of sterile neutrinos,
and then conclude in Sec.~4.

\section{Captures on Beta-decaying or EC-decaying Nuclei}
\label{detection}

In the presence of 3+$N_{\rm s}$ species of active and sterile
neutrinos, the flavor eigenstates of three active neutrinos and $N_{\rm s}$ sterile neutrinos can be
written as~\cite{XZ,PDG}
\begin{eqnarray}
\left(
\begin{matrix}
\nu^{}_e \cr \nu^{}_\mu \cr \nu^{}_\tau \cr \vdots \cr
\end{matrix}
\right)
= \left(\begin{matrix} U^{}_{e1} & U^{}_{e2} & U^{}_{e3} & \cdots \cr
U^{}_{\mu 1} & U^{}_{\mu 2} & U^{}_{\mu 3} & \cdots \cr
U^{}_{\tau 1} & U^{}_{\tau 2} & U^{}_{\tau 3} & \cdots \cr
\vdots & \vdots & \vdots & \ddots \cr \end{matrix} \right)
\left(\begin{matrix} \nu^{}_1 \cr \nu^{}_2 \cr \nu^{}_3 \cr \vdots \cr
\end{matrix}
\right) \; ,
%     (3)
\end{eqnarray}
where $\nu^{}_i$ is a mass eigenstate of active (for $1 \leq i \leq 3$) or sterile (for $4 \leq
i \leq 3 + N_{\rm s}$) neutrinos, and $U^{}_{\alpha i}$ stands for an element of the $(3+ N_{\rm s}) \times (3+ N_{\rm s})$ neutrino mixing matrix.

\begin{figure}[t]
\begin{center}
\begin{tabular}{c}
\includegraphics*[width=0.45\textwidth]{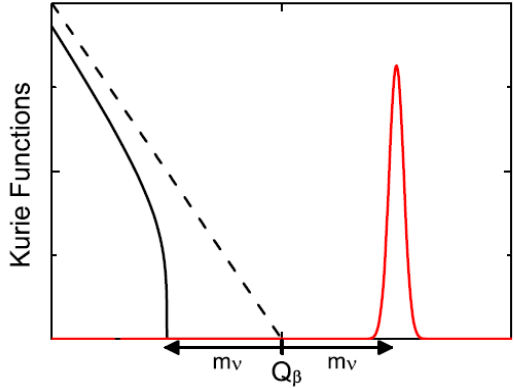}
\end{tabular}
\end{center}
\vspace{-0.8cm}
\caption{Idealized electron spectra for the tritium beta decay and relic neutrino capture. The
dashed and black-solid lines are shown for $\beta$-decay spectra of the massless and massive neutrinos respectively.
The red-solid line with the sharp peak is for the relic neutrino signal.}
\label{capture}
\end{figure}

For the nuclear $\beta$-decay process with the mass number $A$ and atomic number $Z$ of the parent nucleus,
i.e. ${\cal N}(A,Z) \to {\cal  N}^\prime (A, Z+1) + e^- + \overline{\nu}^{}_e$,
the differential decay rate of a $\beta$-decay can be written as \cite{Weinheimer}
\begin{eqnarray}
& &\frac{{\rm d} {\lambda}^{}_\beta}{{\rm d}T^{}_e}  =
\int_0^{Q^{}_{\beta}- {\rm min}(m^{}_i)} {\rm d} T^\prime_e \,
\left\{\frac{G^2_{\rm F} \, \cos^2\theta^{}_{\rm C}}{2\pi^3} \,
F\left(Z, E^{}_{e}\right) \, |{\cal M}|^2 E^{}_{e}\sqrt{E^2_e -
m^2_e}   \,  \right .
\nonumber \\
& & \left . \times\left(Q^{}_{\beta} - T^\prime_e\right)\sum^{3+ N_{\rm s}}_{i=1}
\left[ |U^{}_{ei}|^2\sqrt{\left(Q^{}_{\beta}- T^\prime_e \right)^2 -
m_i^2} ~ \Theta\left(Q^{}_{\beta} - T^\prime_e - m^{}_i\right)
\right]\right\}
%\nonumber \\
\times R\left(T^{}_e, T^\prime_e\right) \; ,
%      (5)
\end{eqnarray}
where $T^\prime_e = E^{}_e - m^{}_e$ denotes the kinetic energy
of the outgoing electron, $F(Z, E^{}_{e})$ is the Fermi function,
$|{\cal M}|^2$ is the dimensionless nuclear matrix elements \cite{Weinheimer},
and $\theta^{}_{\rm C} \simeq 13^\circ$ is the Cabibbo angle.
In addition, a Gaussian energy resolution function %$R(T^{}_{e}, \, T^{\prime}_{e})$
\begin{equation}
R(T^{}_{e}, \, T^{\prime}_{e}) = \frac{1}{\sqrt{2\pi} \,\sigma}
\exp\left[-\frac{(T^{}_{e} - T^{\prime}_{e})^2}{2\sigma^2} \right] \; ,
%     (6)
\end{equation}
is implemented in Eq.~(3) to include the finite energy resolution, and the theta function
$\Theta(Q^{}_{\beta} - T^\prime_e - m^{}_i)$
is adopted to ensure the kinematic requirement. The spectral shape near the $\beta$-decay
endpoint represents a kinetic measurement of the absolute neutrino masses, which can
be understood by comparing the dashed and black solid lines of Fig.~\ref{capture}.
%%%%%%%%%%%%%%%%%%%% Fig. 1 %%%%%%%%%%%%%%%%%%%%%%%%%%%%%%
%%%%%%%%%%%%%%%%%%%%%%%%%%%%%%%%%%%%%%%%%%%%%%%%%%%%%%%%%%

On the other hand, the threshold-less neutrino capture process,
\begin{eqnarray}
{\nu}^{}_e + {\cal N}(A,Z) \to {\cal  N}^\prime (A, Z+1) + e^-\,,
%     (7)
\end{eqnarray}
is located well beyond the end point of the $\beta$-decay, where the signal is characterized by
the monoenergetic kinetic energy of the electron for each mass eigenstate (see the red-solid line in Fig.~1).
This capture process is suitable to detect relic active and sterile neutrinos, and
a measurement of the distance between the decay and capture processes will directly
probe the C$\nu$B. 
The differential neutrino capture rate of this process reads
\begin{equation}
\frac{{\rm d} \lambda^{}_{\nu}}{{\rm d} T^{}_{e}} = \sum_i
|U^{}_{ei}|^2 \sigma^{}_{\nu^{}_i} v^{}_{\nu^{}_i}
n^{}_{\nu^{}_i}\,R(T^{}_{e}, \, T^{\prime i}_{e}) \; ,
%     (8)
\end{equation}
where the sum is for all the neutrino mass eigenstates and
$n^{}_{\nu^{}_i}\equiv\zeta^{}_i \cdot\langle n^{}_{\nu^{}_i}\rangle$
denotes the number density of the relic neutrinos $\nu^{}_i$ around the Earth. The standard Big Bang cosmology
predicts $\langle n^{}_{\nu^{}_i} \rangle \approx \langle n^{}_{\overline{\nu}^{}_i} \rangle \approx 56 ~{\rm cm}^{-3}$
for each species of active neutrinos,
and it is also expected to hold for each sterile neutrino species if they could be fully thermalized in the early Universe.
The number density of relic active and sterile neutrinos may be enhanced by the 
gravitational clustering effect (i.e., the factor $\zeta^{}_i$) when the neutrino mass is greater than 0.1 eV \cite{Wong}.
The capture cross-section times the neutrino velocity can be written as
$\sigma_{\nu_i} v_{\nu_i}={2\pi^{2}\ln2}/(\emph{A}\times T_{1/2})$,
%\begin{eqnarray}
%\sigma_{\nu_i} v_{\nu_i}=\frac{2\pi^{2}}{\emph{A}}\cdot\frac{\ln2
%}{T_{1/2}}\,,
%%     (10)
%\end{eqnarray}
where $\emph{A}$ is the nuclear factor characterized by $Q_{\beta}$ and $Z$ and
$T_{1/2}$ is the half-life of the parent nucleus.

To get a better signal-to-background ratio, one can investigate different kinds of candidate nuclei by considering
factors including the cross-section, half-life, $\beta$-decay rate,
and detector energy resolution. Based on this selection criterion,
several promising nuclei such as $^{3}$H, $^{106}$Ru, and $^{187}$Re are identified after
an exhaustive survey in Ref.~\cite{Cocco}.

The $\beta$-decay experiments of current generation include the spectrometer of KATRIN~\cite{KATRIN} and the calorimeter of MARE~\cite{MARE}.
KATRIN uses 50 $\mu$g of $^{3}$H as the effective target mass, and MARE is planning to deploy 760 grams of $^{187}$Re.
Therefore, we can estimate their respective C$\nu$B event rates to be $10^{-6}\;{\rm yr}^{-1}$ and $10^{-7}\;{\rm yr}^{-1}$
without considering the gravitational clustering effect.
A first realistic proposal for the C$\nu$B detection is the PTOLEMY project~\cite{PTOLEMY}, which is designed to employ 100 grams of $^{3}$H
as the capture target using a combination of a large-area surface-deposition tritium target, the MAC-E filter, the RF tracking,
the time-of-flight systems, and the cryogenic calorimetry. Finally, the event rate of PTOLEMY are calculated to reach the observable level:
\begin{eqnarray}
N^{\nu}({\rm PTOLEMY})\simeq8.0\times\sum_{i}|U^{}_{ei}|^2\zeta^{}_i\quad {\rm
yr}^{-1}\,.
\end{eqnarray}

According to Eq.~(5), only electron neutrinos can be captured in the $\beta$-decaying nuclei.
One should consider other possibilities for the cosmic antineutrino background detection.
Similar to the process of captures on $\beta$-decaying nuclei, the EC-decaying nuclei can be
the target of relic antineutrino captures. The isotope $^{163}{\rm Ho}$ is a promising candidate in this respect~\cite{Cocco2,Lusignoli,LX11EC,LXJCAP}.
The properties of the relic antineutrino capture
against the EC-decaying background are similar to those of $\beta$-decaying nuclei~\cite{LX11EC}. As the order of magnitude estimate,
one needs 30 kg $^{163}{\rm Ho}$ to obtain one event per year for the relic antineutrino detection.

\section{Flavor Effects}
\label{flavor}

%%%%%%%%%%%%%%%%%%%% Fig. 2 %%%%%%%%%%%%%%%%%%%%%%%%%%%%%
\begin{figure}[t]
\begin{center}
\begin{tabular}{cc}
\includegraphics*[bb=18 18 280 216, width=0.40\textwidth]{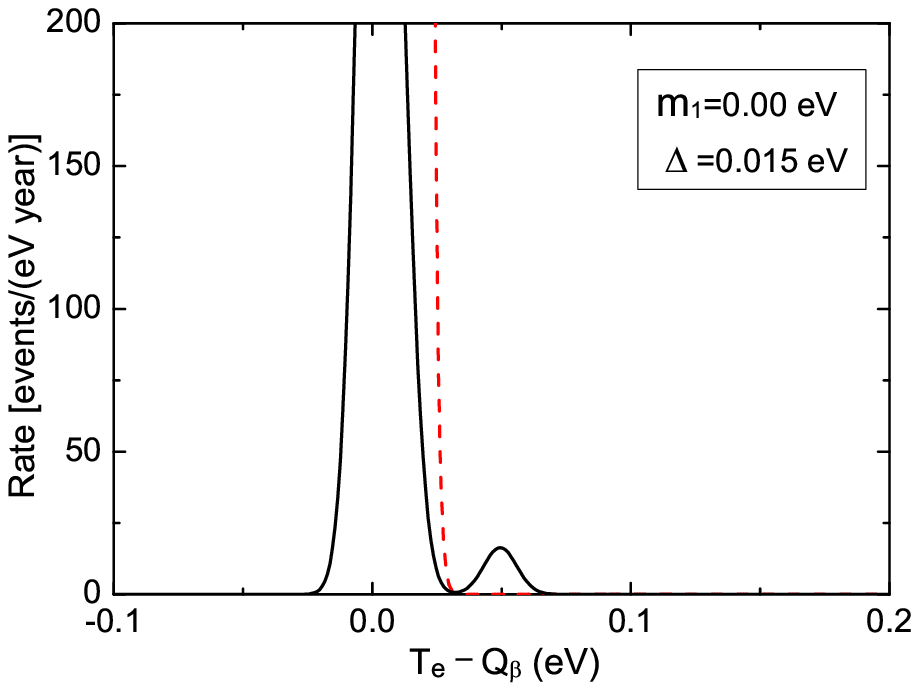}
&
\includegraphics*[bb=18 18 280 216, width=0.40\textwidth]{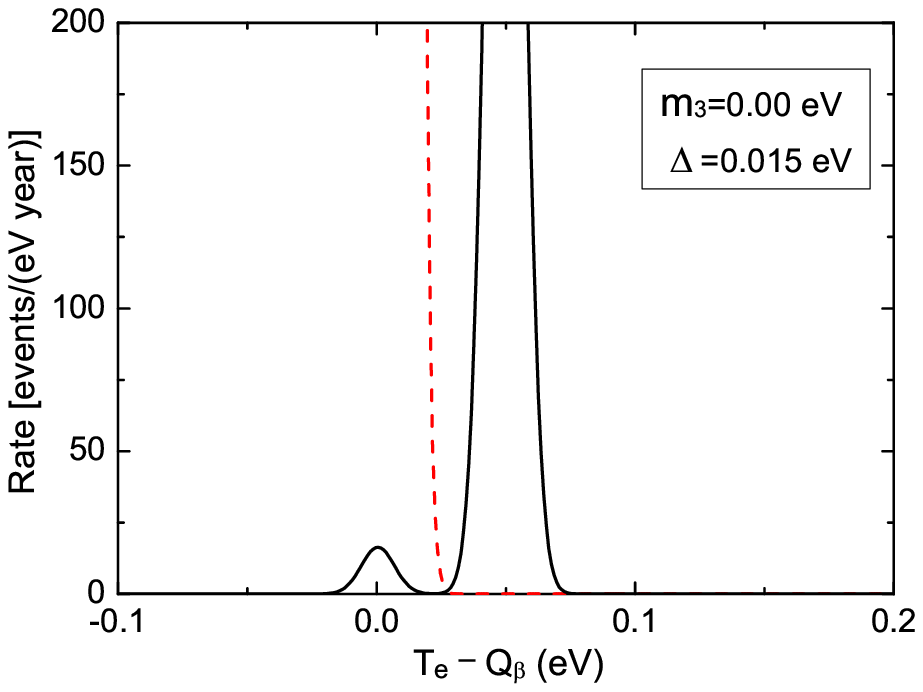}
\\
\includegraphics*[bb=18 18 280 216, width=0.40\textwidth]{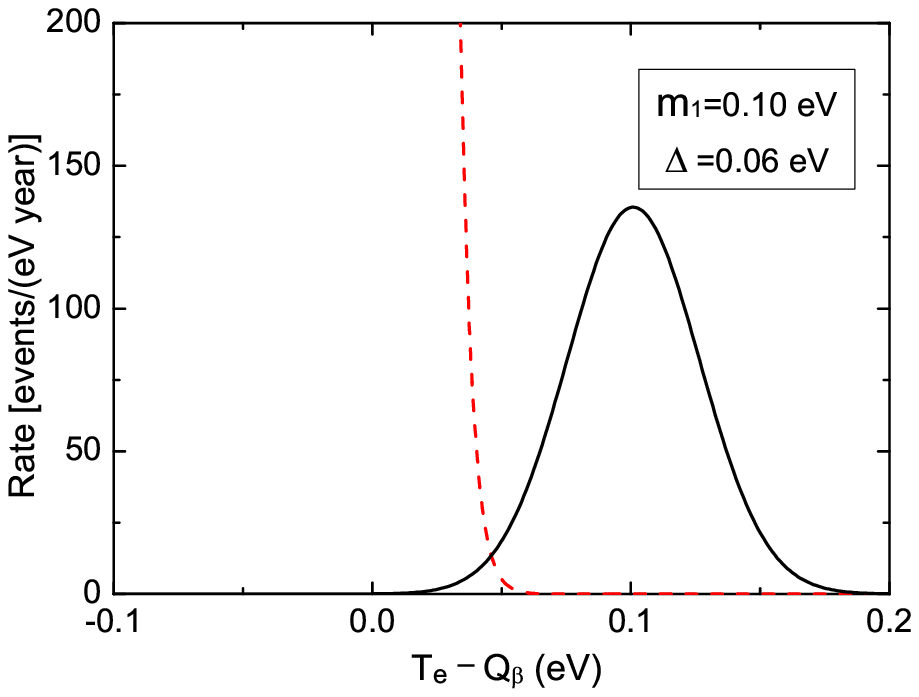}
&
\includegraphics*[bb=18 18 280 216, width=0.40\textwidth]{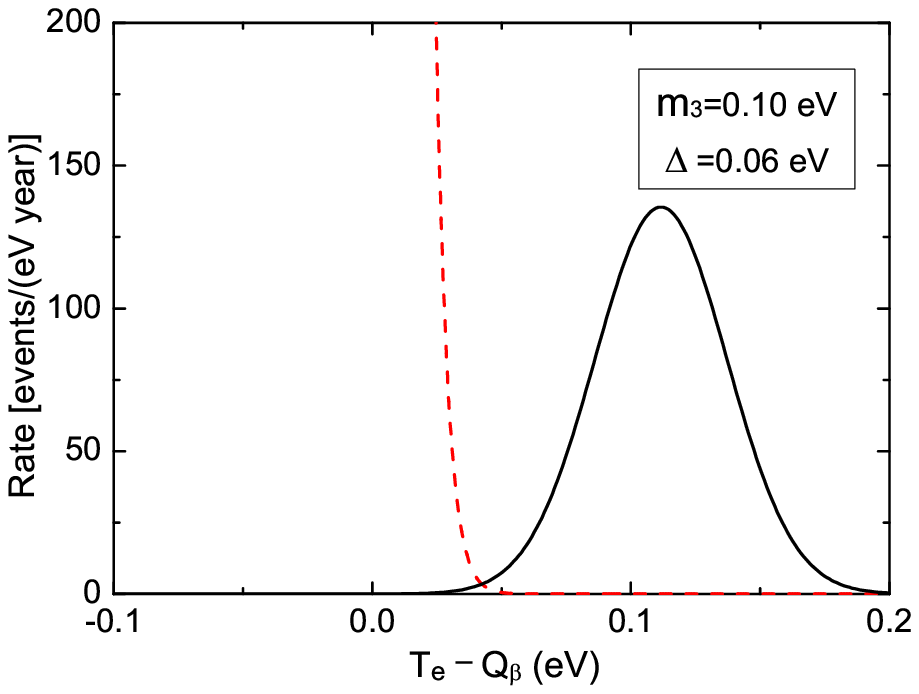}
\end{tabular}
\end{center}
\vspace{-0.6cm}
\caption{The relic neutrino capture rate as a function of the
kinetic energy of electrons in the standard scheme with $\Delta
m^2_{31} > 0$ (left panel) or $\Delta m^2_{31} < 0$ (right panel). The
gravitational clustering of three active neutrinos has been neglected for simplicity.
We adopt 100 grams of $^{3}$H, and best-fit values of the relevant three-neutrino oscillation parameters from Ref.~\cite{PDG}.}
\end{figure}
%%%%%%%%%%%%%%%%%%%%%%%%%%%%%%%%%%%%%%%%%%%%%%%%%%%%%%%%%%
Besides the total capture rates, the C$\nu$B detection exhibits interesting properties of flavor effects
due to the neutrino mixing. In this section, we shall discuss the effects of the neutrino mass hierarchy~\cite{Blennow}
and presence of light sterile neutrinos~\cite{LLX}.

Fig.~2 shows the capture rate of the C$\nu$B as a function of the kinetic energy $T^{}_e$ of electrons in the standard three-neutrino scheme with
$\Delta m^2_{31} > 0$ and $\Delta m^2_{31} < 0$.
The gravitational clustering of three active neutrinos has been neglected for simplicity.
$\Delta$ (i.e., $\Delta = 2\sqrt{2\ln 2} \,\sigma$) denotes the finite energy resolution.
As the lightest neutrino mass increases from 0 to 0.1 eV, the neutrino capture signal moves towards the
larger $T^{}_{e}$ region. The distance between the signal peak and the $\beta$-decay background becomes larger for
a larger lightest neutrino mass, and therefore the required energy resolution is less stringent.
Comparing between the left panel and right panel, one can observe that it is easier to detect the C$\nu$B in the
$\Delta m^2_{31} <0$ case, where the capture signal is separated more apparently from the $\beta$-decay background.
The reason is that the dominant mass eigenstates $\nu^{}_1$ and $\nu^{}_2$ in $\nu_{\rm e}$ have greater eigenvalues than in the $\Delta m^2_{31} > 0$ case.

Next we shall study the (3+2) mixing scheme with two light sterile
neutrinos. Considering the hints of short baseline oscillations \cite{sterile,sterile2},
we assume $m^{}_4 = 0.2$ eV and $m^{}_5 = 0.4$ eV together with
$|U^{}_{e1}| \approx 0.792$, $|U^{}_{e2}| \approx 0.534$,
$|U^{}_{e3}| \approx 0.168$, $|U^{}_{e4}| \approx 0.171$ and
$|U^{}_{e5}| \approx 0.174$ in the numerical calculations.
We illustrate the capture rate of 
the C$\nu$B against the corresponding $\beta$-decay background for both $\Delta
m^2_{31} > 0$ and $\Delta m^2_{31} < 0$ schemes in Fig.~3. To take account of possible gravitational
clustering effects, we assume $\zeta^{}_1 = \zeta^{}_2 = \zeta^{}_3 =1$ and $\zeta^{}_5 = 2\zeta^{}_4 = 10$.
As one can see from Fig.~3, the signals of sterile neutrinos are
obviously enhanced because of $\zeta^{}_4 >1$ and $\zeta^{}_5 >1$. If
the overdensity of relic neutrinos is very significant around the Earth,
it will be helpful for the C$\nu$B detection through the neutrino capture process.
%%%%%%%%%%%%%%%%%%%% Fig. 3 %%%%%%%%%%%%%%%%%%%%%%%%%%%%%%
\begin{figure}[t]
\begin{center}
\begin{tabular}{cc}
\includegraphics*[bb=18 18 274 212, width=0.40\textwidth]{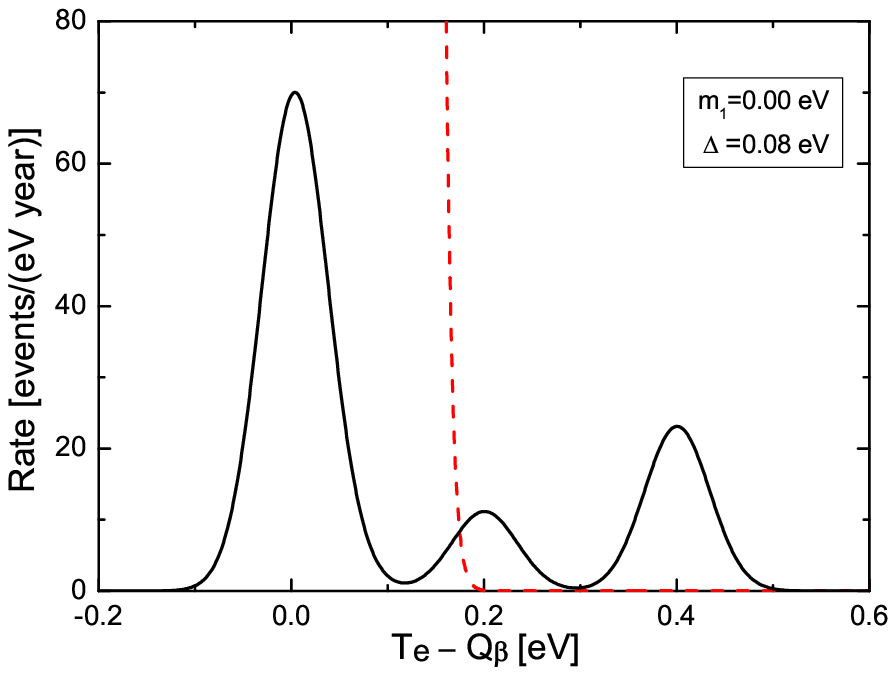}
&
\includegraphics*[bb=18 18 274 212, width=0.40\textwidth]{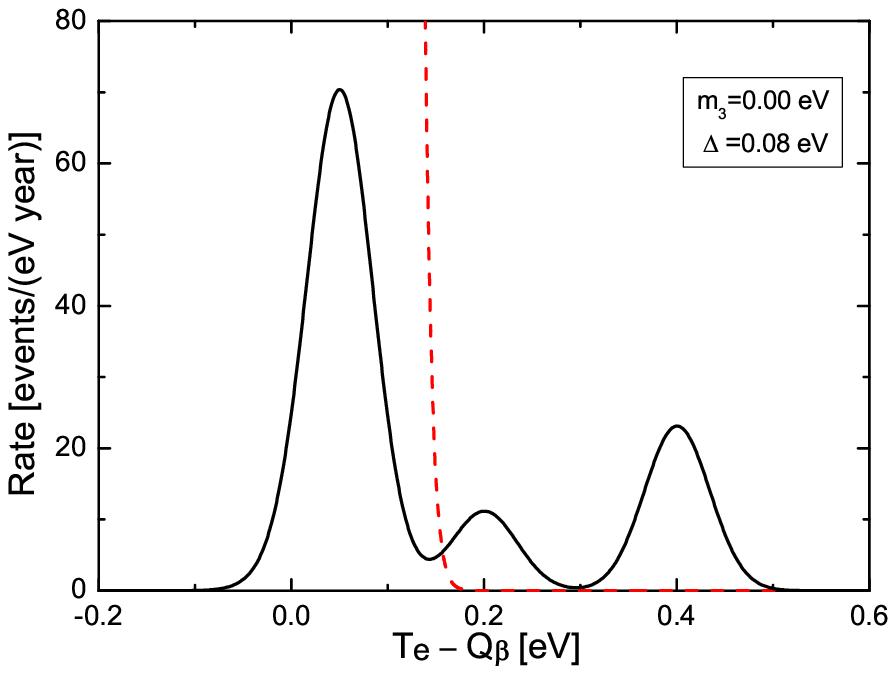}
\end{tabular}
\end{center}
\vspace{-0.6cm}
\caption{The capture rate of the C$\nu$B as a function of the electron's kinetic
energy in the (3+2) mixing scheme with $\Delta m^2_{31} > 0$
(left panel) and $\Delta m^2_{31} < 0$ (right panel) \cite{LLX}. The
gravitational clustering of relic sterile neutrinos around the Earth
has been illustrated by taking $\zeta^{}_1 = \zeta^{}_2 = \zeta^{}_3
=1$ and $\zeta^{}_5 = 2 \zeta^{}_4 = 10$.}
\end{figure}
%%%%%%%%%%%%%%%%%%%%%%%%%%%%%%%%%%%%%%%%%%%%%%%%%%%%%%%%%%

\section{Conclusion}
\label{conc}

The standard Big Bang cosmology predicts the existence of a cosmic neutrino background formed
at an age of one second after the Big Bang. %a temperature of about one MeV and
A direct measurement of the relic neutrinos would open a new window to the early Universe.
We have discussed the future prospect for the direct detection of the C$\nu$B,
with the emphasis on the method of captures on $\beta$-decaying nuclei and PTOLEMY project.
We calculated the neutrino capture rate against the corresponding $\beta$-decay background, and discussed
the possible flavor effects including the neutrino mass hierarchy and presence of light sterile neutrinos.
We stress that such direct measurements of the C$\nu$B in the laboratory experiments might not be
hopeless in the long term.

\section*{Acknowledgement}

This work was supported by the National Natural Science Foundation of China under grant Nos. 11135009 and 11305193,
and the CAS Center for Excellence in Particle Physics (CCEPP).

\end{document}